# On the Ambiguity of Commercial Open Source

**Dr. Ioan Despi, Lecturer**
University of New England, Armidale, Australia
**Dr. Lucian Luca, Associate Professor**
"Tibiscus" University, Timişoara, România

**Abstract**. Open source and commercial applications used to be two separate worlds. The former was the work of amateurs who had little interest in making a profit, while the latter was only profit oriented and was produced by big companies. Nowadays open source is a threat and an opportunity to serious businesses of all kinds, generating good profits while delivering low costs products to customers. The competition between commercial and open source software has impacted the industry and the society as a whole. But in the last years, the markets for commercial and open source software are converging rapidly and it is interesting to resume and discuss the implications of this new paradigm, taking into account arguments pro and against it.

## 1. Introduction

Google, Amazon, Wikipedia, eBay, just to mention the prominent ones, rely all on open source software. Open source is a relatively new development method in software engineering that opposes the classical commercial software development with regard to implied processes and final product. Briefly, openness means that the inner parts of the programs are displayed and anyone can legally access the source code and make changes to it.

The basic idea was that anyone that needed some specific functionality wrote some code, shared it with others, and in return got further enhancements of the functionality. At the very beginning, open source projects were self-organising dispersed communities of programmers who worked together using version control systems (like CVS [BF03]) and Internet to design, code and maintain a particular program. Participants may be volunteering to contribute or some may be affiliated with other organisations (with a declared or undeclared interest in open source) and traditionally responsibilities and tasks were not assigned [MFH02].

Open Source development complements and competes with commercial development. Even so, the current landscape is dominated by commercial software that offers the opposite of OSS/FS - that is, closed or proprietary source software with a price tag. According to a study by consulting firm Optaros (www.optaros.com), 87% of organisations are now using open source software somewhere. This is also a consequence of some major events happened last year [La05]:
- Red Hat, which sells and supports a version of Linux for businesses, had revenues up 43.6%, to $73.1 million, and profits up 114%.
- Sun Microsystems opened sources everything except Java
- Novell bought Suse Linux two years ago and is trying to revive its business through open source.
- Motorola announced that Linux would be its standard operating system for its future phones.
- Firefox celebrated its 100 millionth download in October, just before its first birthday.
- MySQL had $40 million revenues
- About $400 million was invested in open source startups. A good example is SugarCRM which makes software competing with Siebel.

However, the open source approach has moved far beyond its origins to fields like biotechnologies, new drugs, legal research, etc. that could not contain any software at all and the open source buzzword means contributions for various individuals to create something that becomes freely available to everybody.

A good example is given by the journal `Nature`: instead of sending submissions out to a few experts, the journal posts submissions on a site running Movable Type, and then waits for comments. More, the journal is conducting an open debate on peer review (see http://blogs.nature.com/nature/peerreview/trial for more details).

Another example is provided by Toyota, that has decentralised its teams in such a way that they provide the same sort of flexibility and autonomy as Linux communities [EW05].

## 2. The Open Source Concept

Open source is the new catalyst for competition in the 21st century software market. It forces software applications to compete, sell, and survive based on their functionality and quality, not on the marketing excessive publicity. The first open source movement was Richard Stallman's Free Software Foundation, established in 1984 and "dedicated to promoting computer users' rights to use, study, copy, modify, and redistribute computer programs" [www.fsf.org].

The method used was Copyleft and the resulting licence was called GNU General Public Licence (GNU GPL). The aim of FSF would be to support different projects, such as Free Software Licensing and Compliance Lab (www.fsf.org/fsf/licensing), Free Software Directory (http://directory.fsf.org, Savannah (http://savannah.gnu.org), GNU (www.gnu.org), GPLv3 (http://gplv3.fsf.org). GNU (an acronym of *GNU's s Not UNIX*) was established in 1984 to write and maintain software packages, such as GNU/Linux, GNU Emacs, GNU GCC, etc. As it is written in the GNU Manifest, "… we support the freedom of speech, press, and association on the Internet, the right to use encryption software for private communication, and the right to write software unimpeded by private monopolies". In 1997 Bruce Perens wrote the first draft of the Open Source Definition (www.opensource.org) and in 1998 at a Palo Alto strategy session, the term Open Source was coined by the group composed by T Anderson, C Peterson, J Hall, L Augustin, S Ockman, and E Raymond. This coincided with the source code release of the Netscape Navigator (see [OSI] for the entire history). It is largely accepted today that both terms refer to the same concept, but still have different values.

Free Software/Open Source has come a long way in these twenty years from its inception of being just a way of making software available to users to a development method recognised by software engineering discipline. The first and most visible success on the commercial software market was the large acceptance of the all flavours of Linux operating system, which was created inside the GNU project. It was soon followed by other programs, such as Apache web server, the MySQL database management system, languages as Perl, Python and PHP, forming together a structure baptised LAMP. Soon, the companies producing commercial enterprise software realised that they should start using open source projects as modules or components into their own products. Apache became the most used web server, and big companies as Oracle or SAP certified their products for use on Linux. Other big players, as IBM and Dell, began to provide Linux as a pre-installed, formally supported operating system on their PCs and servers.

Last year, Eben Moglen et al. started the process of updating the General Public Licence (http://gplv3.fsf.org) such that it will include issues as patents and online services. The drafting process uses the same approach as the open software production itself, and it expects contributions from every developer, distributor or user and there will be even a conference (the fourth) in August 2006 in Bangalore, India.

## 3. The Open Source Business Model

It is often hard to understand how an open source company may give away its products for free or for a minimal cost and still to survive in the capitalistic market. The answer is that it still generates stable and scalable revenue streams. The open source business model is based on generating revenue from services like systems integration, support, tutorials, and documentation, and not on selling the actual product, which is given away for free. In this way, the emphasis is on the so called "product halo", or ancillary services. The open source approach also cuts down research and development costs, in the same time speeding up the delivery of the new products. An important feature of the open source business model is its ability to market itself.

However, it is unclear how innovative and sustainable is the open source approach. One of the most critical issues is the lack of quality assurance and the handling of intellectual property. If anyone can contribute, than projects are exposed to abuse, being it intentional or generated by ignorance. Projects that fail to obey open source rules or fail to cope with the vulnerabilities of this approach usually give up. Of the about 130000 projects on SourceForge (www.sourceforge.net) only a few hundred are active and even fewer will ever end in a commercially viable product.

Another issue is that rather being a democracy, open source looks like a meritocracy. It is egalitarian at the contribution level but it is elitist when to accept and implement contributions. It seems that the most important obstacle for the open source approach is itself.

The core of the open source movement is community and open source culture is about participation, not profit. The new software development norm is collaboration. The central point to the success of this method is to understand why individuals work "free of charge" in Open Source projects. Before the Internet, it was harder to collaborate on projects or to reach an audience, even if people have always been willing to do good work for free and to work a lot harder on stuff they like. Many researchers have wondered why people give their work away and the short answer is that respect must be earned and cannot be derived from position. The main motivations are learning goals, enhanced social relations and reputation, gaining privileged access to a community, or even for a payslip [KSL02].

Developers have been shown to consist of a very diverse group of people, such as professionals, hobbyists, students and a serious issue is whether the motivation of contributors can be sustained. Another doubt is how innovative open source can remain in the long run, as some authors [We04] believe open source might already have reached a self-limiting state. Many projects incorporated in order to protect themselves from individual liability and so they face new challenges, such as lack of resources, the "take-over" attack from the side of big companies, and the inherent bureaucracy.

## 4. The Impact of Open Source

Traditional companies are good at getting people to wake up at dawn for a day's boring and dull labour but it is unclear for the open source approach how the projects can maintain their momentum. A world in which communications are costly favour people working alongside each other, however a world in which they are (almost) free allow people to be dispersed all over the globe. Both approaches have good and bad consequences. One consequence from the latter is that it permits open source communities to grow and produce open software, which is better than the closed one. Closed software is doing at least three bad things: it creates a false sense of security, the good guys will not find holes and fix them, and when a hole is revealed, it makes harder to distribute trustworthy fixes [Co99]. We present the impact of open source on customers, hackers, and businesses.

The main advantage for a software customer is the fact that he is not the prisoner of the vendor anymore. Having access to the source code, a customer can survive even after the vendor closed his business, not to mention he can fix bugs by himself anytime, without waiting for new releases. More, if the vendor's support fee is too big, there are a lot of other people there interested in offering support for a smaller fee. Also, given that most of the open source software can be freely copied and used, there are no licences to track, and thus no related costs or legal risks.

The impact of open source on hackers was anticipated by Eric S. Raymond in [Ray97] and it can be resumed as in [www.opensource.org/advocacy](www.opensource.org/advocacy) : "Internet and Unix hackers, as a rule, already understand the technical case for open source quite well. It's a central part of our engineering tradition, part of our working method almost by instinct. It's how we made the Internet work." There are a lot of companies (Red Hat, IBM, Zope) making money programming open source software and paying the hackers. In the worst scenario, when all the free sources are out there, probably at most 25% of programmers will loose their jobs, as there is still enough to work on vertical code programming.

The impact of open source on businesses is huge. It changed the way businesses understand and produce standards, security, collaboration, reliability, safety, maintenance, etc. The main advantages over the commercial model of producing software are development speed, lower overhead, closeness to the customer, broader market. From the investor's point of view, there are at least the following models of making money with open source ([www.opensource.org/advocacy](www.opensource.org/advocacy)):

**Support Sellers.** Give away the source code, but sell distribution, branding and after-sale service (Red Hat).

**Loss Leader.** Give away the source code as a loss-leader and market positioner for closed software (Netscape).

**Widget Frosting.** A hardware company goes open source to get better drivers (no profit anyway) and interface tools (Silicon Graphics supports Samba).

**Accessorizing.** Selling accessories (books, complete systems, hardware) with open source preinstalled.

There are many strategies around open source applications that bring competitive advantage for hardware and software vendors. A good taxonomy is given in [Ko04], where John Koening distinguishes between the optimisation, dual licence, consulting, subscription, patronage, hosted, and embedded strategies. His observations are gold mine for managers who try to adopt open source for their companies, and for investors who try to evaluate the companies for including into their portfolios.

## 5. The Hybrid World

For about thirty years, enterprises had to stay in the comfort and safety of licensing traditional closed enterprise software. They ending up by totally depend on these applications to automate vast amounts of their activities, from managing supply chains to accounting and human resources. And, of course, the software companies specialised to supply everything an enterprise can dream for. This marriage produced complaints from both sides. From the market side, customers would like to pay less and not to be locked in to a particular software system. The enterprise software suppliers responded in many different ways to those complaints and in the past few years they used the open source approach to address these and other aspects of traditional closed enterprise software.

Historically, companies like Red Hat and Suse decided to improve upon the limited support that the open community was providing for Linux, thus making their operating systems more attractive to enterprise customers. They sold subscriptions to their own versions of Linux so that, for a price, customers got installation scripts, full maintenance and support, as well as a collection of pre-tested open source applications. The next step is given by traditional companies, acting as commercial software firms but making open source licensing and practices the main strategy in all phases of product's life. The representative example is Jboss, starting out to make an application server and finishing by creating an enterprise-class Java development environment.

A good example of a hybrid business model is MySQL, founded in 1995, which gives away its database management system software under an open source licence. Simultaneously, it is selling the software along with maintenance and support contracts to about 8000 customers who pay between 1% to 10% of the regular amount of a similar proprietary product. MySQL estimates that for every paying customer there are out there 1000 people using its product for free. The company employs about 60 developers, based in 25 countries, of whom 70% work from home. The community of users provides free feedback on new features and old bugs but the company rarely accepts code coming from outside hackers.

One major trend in the last twelve months in the markets for commercial and open source enterprise software was convergence between the open source and commercial entities. IBM bought Gluecode, Oracle bought Sleepycat Software, Red Had bought Jboss, all the latter open source companies. Another trend was massive investments by venture capitalists in startups and established open source companies, as the prognosis is $60 billion enterprise software market in the following years.

The new hybrid approach can be resumed by the following features:
- a simplified but functional version of the software is available as open source
- a full version with advanced features is available for a fee. The customer has access to the source code
- documentation may be free or may be available for a fee/subscription
- community support is available for free and offers enhancements to the open source version of the software

- the phone /email support is available for a fee

As the hybrid model stands in the middle between the old closed software and the new open one, it collects benefits and claims from both sides. The main benefit for customers is the lower price of the final product and for the software house the lower development price. Then they come the benefits of the try-before-you-buy approach, where users are able to download, try and experiment the open source version at no charge. They can play with many similar open source products and to choose the one best suited to their needs. In many cases, they don't have to buy the enhanced and payable version of the software or they don't need to pay the fee for support. Even if the customers are snugly with the possibility to openly see, modify and enhance code, they do not use it too much because of the lack of development skilled.

With the traditional open source projects, the failure to create skills in the programmers' community can cause an entire project to fall by the wayside, as SourceForge proves. With the hybrid model, the company can continue the work and provide any level of support that may be required to continue development.

The main challenge we see for the future of hybrid model is how it will cope with big integrated suites of enterprise software, such as ERP, CRM or SCM, as nowadays the hybrid model offers usually a single one or two integrated applications. Again, the answer will be a function of cost: whether or not the benefits of the integration of the suite will overcome the cost of integrating hybrids with the rest of the enterprise's systems.

Another challenge arise from the fact that the hybrid model is used today only as an alternative to more expensive commercial products. There will be a day when the hybrid model will face to solve new problems from an area that has never before been automated and nobody knows how the model will succeed. Why not start with the commercial version from the beginning? We believe that the hybrid model will be used only as alternative to mature, established products.

**Conclusions**

Open source software is not inherently good just because it is open source; it is just another tool to sometimes better serve customers, but it is not a business panacea. Companies face a greater risk of being sued over licence issues from their commercial software vendors while open source licensing reduces this risk. We predict the hybrid business model will have a growing and significant impact on the enterprise software market in the next years, imposing itself as the facto model.